\documentclass[%
 aip,
 amsmath,amssymb,
 reprint,%
]{revtex4-1}
\usepackage{amsmath}
\usepackage{amssymb}
\usepackage{color}
\usepackage{graphicx}
\usepackage{tabularx}
\usepackage{dcolumn} 
\usepackage{bm}      
\usepackage{array}   
\usepackage{amsfonts}
\usepackage{url}
\usepackage{xr-hyper}
\usepackage[bookmarks=false,colorlinks,citecolor=red]{hyperref}

\usepackage[utf8]{inputenc}
\usepackage[T1]{fontenc}
\usepackage{mathptmx}

\include{MyCommand}

\newcommand{\BR}[1]{{\color{red}{#1}}}

\begin{document}


\title{Gaussian Processes for Finite Size Extrapolation of Many-Body Simulations}

\author{Edgar Josu\' e Landinez Borda}
\email{edgar\_landinez\_borda@brown.edu}
\affiliation{Department of Chemistry, Brown University, Providence, Rhode Island 02912, USA}
\author{Kenneth O. Berard}
\affiliation{Department of Chemistry, Brown University, Providence, Rhode Island 02912, USA}
\author{Annette Lopez}
\affiliation{Department of Physics, Brown University, Providence, Rhode Island 02912, USA}
\author{Brenda Rubenstein}
\email{brenda\_rubenstein@brown.edu}
\affiliation{Department of Chemistry, Brown University, Providence, Rhode Island 02912, USA}

\date{\today}

\begin{abstract}
Key to being able to accurately model the properties of realistic materials is being able to predict their properties in the thermodynamic limit. Nevertheless, because most many-body electronic structure methods scale as a high-order polynomial, or even exponentially, with system size, directly simulating large systems in their thermodynamic limit rapidly becomes computationally intractable. As a result, researchers typically estimate the properties of large systems that approach the thermodynamic limit by extrapolating the properties of smaller, computationally-accessible systems based on relatively simple scaling expressions. In this work, we employ Gaussian processes to more accurately and efficiently extrapolate many-body simulations to their thermodynamic limit. We train our Gaussian processes on Smooth Overlap of Atomic Positions (SOAP) descriptors to extrapolate the energies of one-dimensional hydrogen chains obtained using two high-accuracy many-body methods: Coupled Cluster theory and Auxiliary Field Quantum Monte Carlo (AFQMC). In so doing, we show that Gaussian processes trained on relatively short, 10-30-atom chains can predict the energies of both homogeneous and inhomogeneous hydrogen chains in their thermodynamic limit with sub-milliHartree accuracy. Unlike standard scaling expressions, our GPR-based approach is highly generalizable given representative training data and is not dependent on systems' geometries or dimensionality. This work highlights the potential for machine learning to correct for the finite size effects that routinely complicate the interpretation of finite size many-body simulations.
\end{abstract}

\maketitle

\section{Introduction}
Over the past few decades, \textit{ab initio} electronic structure methods have transformed our ability to design materials by enabling researchers to predict the macroscopic and emergent behavior of solids from a basic knowledge of their constituent atoms. Researchers can now routinely model the electronic and geometric properties of systems ranging from quantum materials to heterogeneous catalysts with -- or very near -- chemical accuracy. However, the accuracy that accompanies many-body electronic structure methods such as Coupled Cluster (CC) theory, Quantum Monte Carlo (QMC), and many-body perturbation theories often comes at a steep cost: these methods typically scale as a high degree polynomial with system size. For example, Coupled Cluster Singles, Doubles, and Perturbative Triples [CCSD(T)] conventionally scales as $O(N^{3}M^{4})$, where $N$ is the number of electrons and $M$ is the size of the basis set, while Auxiliary Field Quantum Monte Carlo (AFQMC) typically scales as $O(N^{2}M^{2}+M^{2}N)$.\cite{HC1} In contrast, mean field methods such as Density Functional Theory (DFT) scale as $O(N^2\log N)$\cite{kresse,goedecker} or $O(N)$,\cite{mohr} when locality is a good approximation, but are only predictive when the degree of electron correlation is mild. Historically, the comparatively steep scaling of many-body methods has thwarted their direct application to solids with large unit and/or supercells, limiting their use to systems with just tens to, potentially, hundreds of atoms. However, such smaller, more computationally-accessible systems cannot manifest the same long-range correlations as are present in larger, more realistic solids, and can exhibit spurious boundary effects that confound their interpretation. Indeed, given the remarkable accuracy of many modern electronic structure methods, these finite size errors are often the largest sources of error in many calculations of solids.\cite{miguel,drummond} This leads to a long-standing conundrum: \textit{if many-body methods may only be directly applied to smaller, finite systems, how can they be leveraged to predict the properties of larger, more realistic solids?}

To increase the feasibility of many-body methods for the prediction of the properties of solids in their infinite-size, ``thermodynamic limit" researchers have developed approaches that correct results for smaller systems to predict the properties of larger systems. 
Such so-called finite size corrections consist of two main contributions: one-body and two-body corrections, which ameliorate the one- and two-body contributions to the total energy, respectively. One-body finite size errors typically stem from shell-filling effects that lead to a mis-estimation of the kinetic energy\cite{drummond,foulkes} and can therefore be corrected by a judicious averaging over k-points.\cite{Mihm_JCTC_2021} For example, in mean field theories, integrating over many points in the first Brillouin zone can be circumvented by instead approximating quantities using mean-value points known as Balderschi points.\cite{baldereschi} While many-body methods such as QMC methods need to integrate over the full simulation supercell, not just one point, twist averaging\cite{twisted} provides a means of averaging over a set of angles (offset vectors) on the Brillouin zone of the supercell that results in a rapid convergence of the one-body effects.\cite{twisted,drummond} In contrast, two-body finite size effects stem from errors in the Coulomb and exchange-correlation interactions and are more challenging to correct. These effects can be alleviated by introducing modified versions of these interactions, such as model periodic Coulomb corrections.\cite{drummond,foulkes} An alternative approach for correcting both one- and two-body finite size effects is to determine finite size corrections using methods that scale more gracefully with system size, such as Density Functional Theory (DFT).\cite{drummond,foulkes,mitas} Such methods are used to estimate the differences in energies between smaller and larger systems, and then these differences are added to the smaller-sized many-body calculations. One such DFT-based approach is the Kwee, Zhang, Krakauer correction (KZK).\cite{kzk} While such corrections are now widely applied to materials, they inherently lack the accuracy that would be possible if many-body corrections that take strong correlation into account were applied. 


Even though these one- and two-body corrections markedly reduce finite-size errors, extrapolations to the thermodynamic limit are often still made to reduce any remaining errors. The simplest approach for performing these extrapolations is to fit many-body results obtained at smaller system sizes (e.g., 2x2 or 3x3 supercells) to functional forms that enable extrapolation to larger system sizes.\cite{mitas,foulkes} Nevertheless, it is often unclear which functional form should be employed since it can vary with the geometry, dimensionality, and electronic phase of the material.\cite{miguel,Mihm_JCTC_2021} This is especially true for systems with atypical geometries and boundary conditions. It also of particular importance for calculations involving excited states, including gap and exciton binding energy calculations, because excited states can be more difficult to converge to their thermodynamic limit.\cite{Gorelov_2023_carbon, gorelov2023electronic,PhysRevA.81.013611} 
When a system's correlation energy converges slowly, the number of points necessary for accurate fitting can exceed computational constraints, limiting the overall utility of such extrapolations and the results they yield.\cite{gruneis,Mihm_JCTC_2021}  

One potentially promising approach for estimating many-body corrections that can reduce this computational expense is machine learning. Machine learning methods surrogate more complex models with regressions that have lower computational complexity, thereby accelerating prediction.\cite{metha,carleo,Nyshadham2019} In the context of condensed matter physics, machine learning has been employed to accelerate the prediction and discovery of new materials based upon the properties of known materials\cite{mlcondmat} as well as to learn the presence of certain phases based upon their known signatures.\cite{mlmetalinsul,carrasqu} Machine learning techniques have moreover recently been harnessed to accelerate and improve the accuracy of quantum Monte Carlo methods (see a more detailed discussion in Section \ref{background}).\cite{Ferminet,paulinet,Booth_PRX_2020,Rath_JCP_2020,Mihm_JCTC_2021,Niu_PRL_2023,carleo,WuCarleo_PRR_2023}  
To approach the problem of determining accurate, many-body finite size corrections, one can analogously imagine using data from smaller system sizes to train machine learning algorithms to predict the properties of systems of larger sizes. An early such work used energies and densities from the density matrix renormalization group to learn the DFT kinetic energy functional of hydrogen chains in the thermodynamic limit.\cite{PhysRevB.94.245129} More recently, while this work was being prepared, Gaussian process regression techniques were shown to be able to successfully learn corrections to coupled cluster calculations in k-space. More specifically, Mihm \textit{et al.} \cite{MLFinite_CC, Mihm_JCTC_2021} employed the transfer structure factor to quantify the finite size effects present in coupled cluster theories' correlation energy. They then innovatively bypassed directly computing the structure factor for G values approaching zero (i.e., in the thermodynamic limit) by flexibly representing the structure factor using Gaussian Process Regression.

In this work, we leverage Gaussian Process Regression (GPR)\cite{gprmw} to learn finite size corrections in real-space to homogeneous (one-dimensional) and inhomogeneous (two-dimensional) hydrogen chains modeled using the first-principles, many-body methods Coupled Cluster (CC) Theory and Auxiliary Field Quantum Monte Carlo (AFQMC). 
Kernel methods like Gaussian processes\cite{gprmw} are advantageous because they are not parametric and make use of Bayesian inference that can come at a lower $O(N_{t}^3)$ (where $N_{t}$ is the size of the training set) cost than more complicated parametric methods such as neural networks that scale with the number of layers employed.\cite{metha} Gaussian processes have also been shown to make equally, if not more, accurate predictions than neural networks when less training data is available, which is an important consideration when training is to be performed on data generated using relatively expensive electronic structure calculations.\cite{NNvsGPR} 
We use Gaussian processes to first predict the energies of one-dimensional, homogeneous hydrogen chains of varying lengths using atomic environment descriptors that enable us to incorporate information regarding the geometry and electronic density of each atom and its neighbors. Importantly, even though machine learning methods are most accurate for interpolation, we demonstrate that training our models on the energies of one-dimensional hydrogen chains containing 10-30 atoms enables us to predict (extrapolate) the energies of chains of more than 100 atoms, nearing the thermodynamic limit, with sub-milliHartree accuracy. To contextualize the accuracy of our methods, we compare the accuracy of our predictions to that of polynomial fits to larger-sized systems, the so-called ``subtraction trick,''\cite{subtrick} and other alternative regression methods. Finally, to demonstrate the generalizability and robustness of our approach, we  show that our technique can readily be adapted to also extrapolate the energies of heterogeneous chains of hydrogen dimers, which possess more free parameters, to their thermodynamic limit. This work thus illustrates that machine learning is a relatively cheap, yet accurate means of correcting for finite size effects in many-body simulations that can potentially address many of the challenges the many-body modeling community faces predicting the properties of solids in the thermodynamic limit.     

In the spirit of a \textit{Faraday Discussion}, in Section \ref{background}, we begin with a discussion of the emerging synergies between machine learning techniques and stochastic electronic structure methods. We then describe the machine learning methods, descriptors, and electronic structure techniques we employ in our finite-size extrapolation research in Section \ref{methods}. We next present our primary results demonstrating our technique's ability to accurately correct for finite size errors in Section \ref{results}. 
We conclude by discussing the relative merits and potential applications of our algorithm in Sections \ref{discussion} and \ref{conclusion}. 

\section{Machine Learning in Stochastic Electronic Structure \label{background}} 

Over the past decade, an increasing amount of research has shown that stochastic electronic structure and machine learning methods can form a very fruitful partnership that both accelerates and extends the capabilities of stochastic methods. Because stochastic electronic structure methods are often more expensive than other common electronic structure methods such as Density Functional Theory, machine learning techniques hold the promise of making stochastic electronic structure techniques less costly. At the same time, the high accuracy of most stochastic electronic structure techniques like Diffusion,\cite{10.1063/1.4822960,10.1119/1.4890824} Full Configuration Interaction,\cite{10.1063/1.469756} and Auxiliary Field\cite{PhysRevLett.90.136401,doi:10.1021/acs.jctc.2c00802} Quantum Monte Carlo methods can provide ML techniques with high-quality data that can be used to correct less accurate predictions.

One triumph of the union of these techniques has been the generation of machine learned force fields from QMC energies and gradients.\cite{doi:10.1021/acs.jpca.2c05904,10.1063/5.0052266,Niu_PRL_2023,ryczko_machine_2022} QMC energies and forces calculated for representative configurations are used to train a variety of different neural networks, e.g., Behler-Parrinello Neural Networks,\cite{PhysRevLett.98.146401} or other architectures, which are in turn used to predict the energies and forces for other configurations, accelerating geometry relaxation and/or \textit{ab initio} molecular dynamics simulations.\cite{Ren2023,Scherbela2022} For instance, in recent work, Diffusion Monte Carlo energies and forces were used to generate a force field using a hierarchical $\Delta$-machine learning scheme based upon the Deep Potential Molecular Dynamics (DPMD) framework\cite{DPMD} that was able to successfully uncover a new phase of hydrogen.\cite{Niu_PRL_2023} Since QMC has historically met challenges calculating forces,\footnote{Without corrections, QMC methods typically possess an infinite statistical variance,\cite{PhysRevE.93.033303} and even with corrections, can scale quite steeply (as $Z_{eff}^{6.5}$ or greater, where $Z_{eff}$ is the effective nuclear charge).} recent work has also exploited machine learning architectures to learn force fields from \textit{energy data alone}.\cite{doi:10.1021/acs.jpca.2c05904} Other ways to further reduce the cost of QMC data generation for training itself employ either $\Delta$-ML\cite{doi:10.1021/acs.jctc.2c01058} or transfer learning.\cite{Broecker2017} These techniques first learn potentials and forces, using data from less accurate, but less costly theories and then correct those force fields by either adding a machine learned correction or  updating the less accurate force field with select higher accuracy information. Both methods capitalize on the fact that less accurate theories can often reproduce much of the correct physical behavior of a system, meaning that high accuracy methods are effectively only needed to correct specific phenomena or regions of the potential energy surface. Further opportunities lie in better harnessing the statistical nature of stochastic methods to more efficiently train such force fields.\cite{Ceperley2023arXiv} Overall, QMC-quality force fields open up the grand possibilities of studying dynamics in large molecular or solid state systems with relatively little overhead, making QMC dynamics a practical reality.

Stochastic methods and machine learning techniques have also been fruitfully paired to develop new neural network-based variational ansatze. The Variational Principle, which states that the ground state wave function of a system can best be approximated by varying the parameters and forms of trial wave functions to minimize the energy of the system, has long been used to produce wave function ansatze in computational quantum chemistry and physics. Often, such ansatze have been optimized using QMC (i.e., Variational Monte Carlo methods) and used either on their own or as starting points for projection-based QMC techniques.\cite{drummond,hoggan_chapter_2016} Historically, the forms of these variational ansatze have been specified based upon knowledge of the chemistry/physics they ultimately aim to describe (e.g., Gutzwiller\cite{10.1143/PTP.30.275} or pairing\cite{PhysRevLett.96.130201} wave functions) or confidence that their form is generalizable and expressive enough to describe the phenomenon under study (e.g., backflow wave functions).\cite{PhysRevLett.122.226401}) Specifying the forms of trial wave functions based upon the physics expected can potentially lead to circular logic in which the physics that is expected to be seen is incorporated into a variational wave function form that then recovers that physics. 

Recently, machine learning has been employed to overcome this limitation by providing a means of creating highly expressive variational wave functions. One of the most popular means of achieving this has been to use deep neural networks to specify a given variational wave function and then to optimize that neural network using the energy and/or variance as a loss function.\cite{hermann_ab_2023} Examples of such variational neural networks include DeepQMC,\cite{DeepQMC} FermiNet\cite{Ferminet}, and PauliNet,\cite{paulinet} all of which have shown promise determining the ground states of challenging chemical systems. PauliNet and FermiNet, for example, use deep neural networks to learn a parameterized form of the Jastrow factor and backflow functions and maintain antisymmetry using Slater determinants. Unlike traditional methods that use single-particle orbitals, FermiNet employs functions invariant under two-electron permutations and incorporates back-flow-like transformations for enhanced accuracy.\cite{ye2024widetildeon2}

An alternative approach to combining the expression and optimization of wave functions with machine learning has been neural network quantum states.\cite{Carleo_Science} One promising form of neural network quantum states established by Carleo and Troyer are Restricted Boltzmann Machines, which implement a representation of the wave function through hidden and visible layers.\cite{Carleo_Science} The Boltzmann distribution models the probabilities associated with different configurations of visible and hidden nodes based on the energy; lower energies are favored to accommodate the variational principle which guides the optimization of wave function parameters. These wave functions can then be extrapolated to larger systems by reusing the learned features of the wave function to initialize a machine learning model applied to a similar, but larger system.\cite{pescia2023messagepassing,PhysRevE.101.053301} This process of transferring the learning done for one type of problem to a related, but different problem makes seemingly out-of-reach problems, such as the thermodynamic limit, computationally feasible. 
Success with the transverse-field Ising model,\cite{Carleo_Science} Heisenberg model,\cite{Carleo_Science} and molecules\cite{CarleoandChoo} has been demonstrated. Akin to the use of GPR in this work, Gaussian Processes have also been used to specify wave functions called Gaussian Process States.\cite{Booth} These wave functions are expressed as the exponential of a GP estimator and thus, as Gaussian processes more generally, are highly generalizable and can provide critical information about uncertainties. Such machine learning-based wave functions offer a potential means of achieving unprecedented levels of accuracy without the need for typically more expensive projection techniques. 

Given these successes combining stochastic methods with machine learning approaches - and the many more we have not been able to discuss due to space constraints - here, we focus on the possibility of using machine learning methods to extend QMC's capabilities in a different way: by facilitating the extrapolation of QMC results to the thermodynamic limit.

\section{Methods \label{methods}} 
\subsection{Gaussian Approximation Potentials (GAP)}\label{gaussian-approximation-potentials-gap-and-kernel-methods}

In this work, we employ Gaussian Process Regression (GPR) to predict finite size corrections for discrete hydrogen chains. We have focused on GPR\cite{gprchem} because it has previously been shown to yield high accuracy results with less training data than comparable methods.\cite{NNvsGPR} This is an especially desirable property when one is interested in performing regressions on data obtained from comparatively costly many-body simulations, since computational expense practically limits how much reference data can reasonably be collected. The Bayesian nature of GPR also makes it possible to compute the variance of its predictions, which greatly facilitates the interpretation of its results.\cite{mackaybook} For these reasons, we employ a GPR-based approach which is very similar in flavor to the Gaussian Approximation Potential (GAP) approach.\cite{gap} We first summarize our approach at a high level and then provide more details in subsequent subsections. 

A GPR is a random process which takes input vectors \(\mathbf{x_{i}}\)  and maps them to random variables \(y=f(\mathbf{x})\) with a multivariate, normal joint distribution with covariance ($K$) \cite{gprmw,mackaybook}

\begin{equation}
p(f(x))\sim\mathit{N}(\mu,K).
\end{equation}
The target function \(f(\mathbf{x_{i}})\) (which yields the energy in this work) is characterized by the expectation value of the distribution $\mu=\left\langle f(\mathbf{x_{i}}) \right\rangle$. Like GAP, we use atomic environment descriptors\cite{atomenv} as input features, \(\mathbf{x_{i}}\) (which are vectors containing the atomic descriptors of a structure $i$). These capture the main features of the electron density of an atom and its neighborhood (its atomic environment) to represent the electronic characteristics of the atoms. The covariance determines how the features are correlated and is specified by the kernel function. In kernel methods such as GPR,\cite{gprmw,rupp} input features \(\mathbf{x_{i}}\) are mapped to a nonlinear, high-dimensional space through the function \(\phi(\mathbf{x_{i}})\). Correlations between descriptors that represent different atomic structures are subsequently represented by taking their inner product in this nonlinear space to yield the kernel   

\begin{equation}
K(x_{i},x_{j})=\phi(\mathbf{x_{i}})\cdot \phi(\mathbf{x_{j}}). 
\end{equation}
Nevertheless, the kernel can be defined in a more arbitrary way as long as it satisfies the properties of a covariance matrix.\cite{rupp} In order to make predictions, Bayesian inference can be used to compute new values of the target function.\cite{gprmw,mackaybook} This is done by extending the distribution to unobserved data, \(\mathbf{y}^{*}\). The idea is to generate a distribution based on the observed data \((\mathbf{y,X})\) using unseen data \(\mathbf{X}^{*}\) to generate the prediction \(\mathbf{y}^{*}\) with the corresponding joint distribution:

\begin{equation}
\begin{bmatrix} \mathbf{y} \\ \mathbf{y}^{*} \end{bmatrix} \sim \mathit{N} \left(  \begin{bmatrix}\mu \\ \mu^{*} \end{bmatrix} \begin{bmatrix} K & K_{*} \\ K_{*}^{T} & K_{**} \end{bmatrix} \right),
\end{equation}
where \(\mu\) and \(\mu_{*}\) denote the means over the training and unobserved data, respectively, and  \(K\), \(K_{*}\), and \(K_{**}\) represent the covariances among the training data, training and unobserved data, and unobserved data, respectively. Based upon Bayes' Rule, the posterior
distribution is Gaussian since the joint distribution is Gaussian. The posterior distribution can be expressed as

\begin{equation}
P(\mathbf{y}^{*}|\mathbf{y}) \sim \mathit{N}(\hat{y},\hat{K}),
\end{equation}
while the predicted mean and variance for an unobserved point may be expressed as

\begin{equation}
\mathbf{y}^{*}=\hat{y}=K^{-1}\mathbf{y}K(\mathbf{X},\mathbf{X}^{*})
\end{equation}
and
\begin{equation}
\boldsymbol{\sigma^{*}} = K(\mathbf{X}^{*},\mathbf{X}^{*})-K(\mathbf{X},\mathbf{X}^{*})^{T}K(\mathbf{X},\mathbf{X})^{-1}K(\mathbf{X},\mathbf{X}^{*}). 
\end{equation}
The functional form of the prediction is equivalent to that produced by Kernel Ridge
regression,\cite{rupp} and can be written in the same way

\begin{equation}
\mathbf{y}^{*}=\hat{y}=(K+\sigma^2)^{-1}\mathbf{y}K(\mathbf{X},\mathbf{X}^{*}).
\label{eq:gppr}
\end{equation}
In this equation, the weights, \(\boldsymbol{\alpha}\),\cite{rupp} are given by

\begin{equation}
\boldsymbol{\alpha}=(K+\sigma^2 I)^{-1}\mathbf{y}.
\label{eq:coeffs}
\end{equation}
Equation \ref{eq:gppr} can be written in terms of the coefficients given by Equation \ref{eq:coeffs}

\begin{equation}
y^{*}= \sum_{i}\alpha_i\cdot K(\mathbf{x_i,x^{*}}),
\end{equation}
where the \(\alpha_i\) are vectors of the coefficients obtained from the regression and \(K(\mathbf{x_i},\mathbf{x^{*}})\) is the kernel between the unseen data, \(\mathbf{x^{*}}\), and the training data, \(\mathbf{x_{i}}\). Kernel methods such as GPR can thus be used to predict the total energy, $E^{*}_{total}$, using the equation

\begin{equation}
E^{*}_{total}=  \sum_{i}\alpha_i\cdot K(\mathbf{x_i,x^{*}}).
\end{equation}
The kernels can be tuned to optimize the prediction of the Gaussian process through the selection of their free parameters, known as hyper-parameters. The most common method of optimizing the posterior is the log-likelihood maximization method. In this work, we use three-way hold-out and log-likelihood maximization over the hyper-parameters.\\

\subsection{Atomic environment descriptors and regression model}

In contrast with physics-based approaches for describing a system, machine learning models are often more expressive, meaning that a single model has the potential to describe many different systems. One way to constrain the predictions of a machine learning model is to include prior physical information in the surrogate model.
This can be achieved by making the model invariant to symmetries, including translational, rotational, or permutation symmetries, or constraints, in order to suppress spurious correlations. These symmetries or constraints are usually incorporated into the model in two ways: explicitly integrating these symmetries into the regression algorithm or designing features that are invariant to the symmetry transformations. 

Here, we incorporate symmetries via the latter approach using Smooth Overlap of Atomic Positions (SOAP) descriptors that are invariant to rotation and translation. These atomic environment descriptors represent the electron density at some point $r$ by the superposition of the Gaussian densities of atoms with the same atomic number \(Z\) in the neighborhood of that point

\begin{equation}
 \rho^{Z}(r)=\sum_{i}^{\|Z_{i}\|}\exp\left(-\frac{\|r-R_{i}\|^2}{2\sigma^2}\right),
\end{equation}
where $R_{i}$ is the position of an atom, $i$, in the neighborhood and $\sigma^2$ is the variance of the Gaussian. This density may be expanded in terms of radial and angular basis functions
\begin{equation}
\label{eq:rho}
 \rho^{Z}(r)=\sum_{nlm} c^{Z}_{nlm}Y_{lm} g_{n}(r),
 \end{equation}
where the \(g_{n}(r)\) are the $n$ radial basis functions that can be expressed in terms of polynomials or atomic orbitals and the \(Y_{lm}\) correspond to the spherical harmonic functions. The $c^{Z}_{nlm}$ coefficients of the expansion can be computed by integrating over the density

\begin{equation}
 c^Z_{nlm}(\mathbf{r}) =\iiint_{\mathcal{R}^3}\mathrm{d}V g_{n}(r)Y_{lm}(\theta, \phi)\rho^Z(\mathbf{r}).
 \end{equation}
 
In this work, we use the Dscribe library \cite{dscribe} to obtain the descriptors. This library implements SOAP descriptors using a partial power spectrum that only includes real spherical harmonics. Because the density depends on the square of the distances between points, it is already invariant to translation. A descriptor vector, $\mathbf{p}$, is formed from elements of the power spectrum

\begin{equation}
p(\mathbf{r})^{Z_1 Z_2}_{n n' l} = \pi \sqrt{\frac{8}{2l+1}}\sum_m c^{Z_1}_{n l m}(\mathbf{r})^*c^{Z_2}_{n' l m}(\mathbf{r}),
\end{equation}
where \(n\) and \(n'\) \(\leq n_{max}\) run over the radial basis functions and \(l\leq l_{max}\) runs over the spherical harmonics. $n_{max}$ and $l_{max}$ define the maximum number of radial and angular functions in which the density in Equation \ref{eq:rho} is expanded, respectively. \(Z_1\) and \(Z_2\) are the atomic numbers of the species. The resulting power spectra are rotationally- and permutationally-invariant by construction.  

The original SOAP descriptors compare the local atomic environments using a kernel that is the dot product of the normalized power spectra between different configurations

\begin{equation}
K^\mathrm{SOAP}(\mathbf{p}, \mathbf{p'}) = \left( \frac{\mathbf{p} \cdot \mathbf{p'}}{\sqrt{(\mathbf{p} \cdot \mathbf{p})\times(\mathbf{p'} \cdot \mathbf{p'})}}\right)^{\xi}.
\end{equation}
This kernel takes the overlap of two atomic environments. However, other kernels employ different ways of measuring the similarity of the environments that may lead to better results.  
One of the most common kernels because of its versatility and robustness is the Radial Basis Function (RBF) or Squared Exponential (SE) kernel

\begin{equation}
K(\mathbf{p},\mathbf{p'})=v^{2}\exp\left(\frac{d(\mathbf{p},\mathbf{p'})^2}{l^2}\right),
\end{equation}
where \(d(\mathbf{p},\mathbf{p'})\) is the Euclidean distance, $v^2$ is a tunable amplitude, and $l^{2}$ is the global weight or length scale of the features. We choose to use the latter kernel throughout this work because of its flexibility and robustness for comparing features.

\subsection{Comparing environments with global
descriptors}\label{comparing-environments-with-local-and-global-descriptors}

The descriptor vector, $\mathbf{p}$, of an atomic structure depends on the number of atoms of each species and is created by concatenating the different combinations of atomic species, each with $n$ radial basis functions and a maximum angular number $l_{max}$.\cite{atomenv} As a result, structures with different numbers of atoms, $M$, $N$, have different numbers of descriptors. One way to deal with descriptor vectors of differing lengths is to pad the feature vectors with zeros such that their dimensions match those of the descriptor vectors with the largest number of features in the samples. A similar approach involves padding the dummy (missing) features with values selected to decrease the biases the missing features would otherwise introduce.\cite{kriggingv}\\

An alternative that can reduce bias is the use of global descriptors. These descriptors characterize the whole structure, i.e., the features depend on all of the atoms, rendering the number of features independent of the number of atoms in the structure. However, such an approach may diminish the quality of the kernel, since the descriptors may not have enough resolution to distinguish subtle differences between structures because of their global nature. A very simple and intuitive method to make the kernel global is to construct an ``average kernel:"\cite{ceriotti}

\begin{equation}
K(A,B)=\frac{1}{NM}\sum_{i,j}^{N,M}C_{ij}(A,B).
\end{equation}
Such a kernel recursively compares the features of the atoms $i$ and $j$ in structures $A$ and $B$, respectively, using the kernel, $C$, and averaging over its corresponding numbers of atoms $N$ and $M$. This approach is equivalent to averaging the features of all of the atoms of each configuration and comparing them with the kernel $C$, which amounts to making the descriptors global

\begin{equation}
\bar{p}(\mathbf{r})^{Z_1 Z_2}_{n n' l} = \pi \sqrt{\frac{8}{2l+1}}\sum_i^{N} \frac{1}{N} \sum_m \left(c^{i,Z_1}_{n l m}(\mathbf{r})^* \right) \left(c^{i,Z_2}_{n' l m}(\mathbf{r})\right).
\end{equation}
The RBF kernel with the global descriptors then becomes 

\begin{equation}
K(\bar{p},\bar{p}^{'})=v^{2}\exp\left(\sum_{i}\frac{d(\bar{p_{i}},\bar{p_{i}}^{'})^2}{l^{2}_{i}}\right).
\label{eq:rbf}
\end{equation}
It is important to note that, when global descriptors are employed, the total energy is no longer the simple sum of local contributions. Now, it explicitly depends on quantities that interrelate features of the whole structure. This overall description of atomic structures implicitly removes the need for descriptors that capture long-range order. Nonetheless, the resolution of the features still needs to be high enough to capture small structural changes, as mentioned earlier. The resolution of the kernel can be improved by weighting each global feature by some characteristic length, $l_i$, according to Equation \ref{eq:rbf}. This improves kernel performance by allowing fine-tuning of the parameters, but at the cost of adding more complexity to the model. In the following, we employ this combination of SOAP-averaged descriptors and the RBF kernel on linear hydrogen chains, which serve as an interesting and challenging benchmark.\\ 

\section{Results \label{results}}

\subsection{One-Dimensional, Homogeneous Hydrogen Chains}

\subsubsection{Coupled Cluster and AFQMC Database of Homogenous Hydrogen Chain Energies} 

\begin{figure}[h!]
\includegraphics[width=0.48\textwidth]{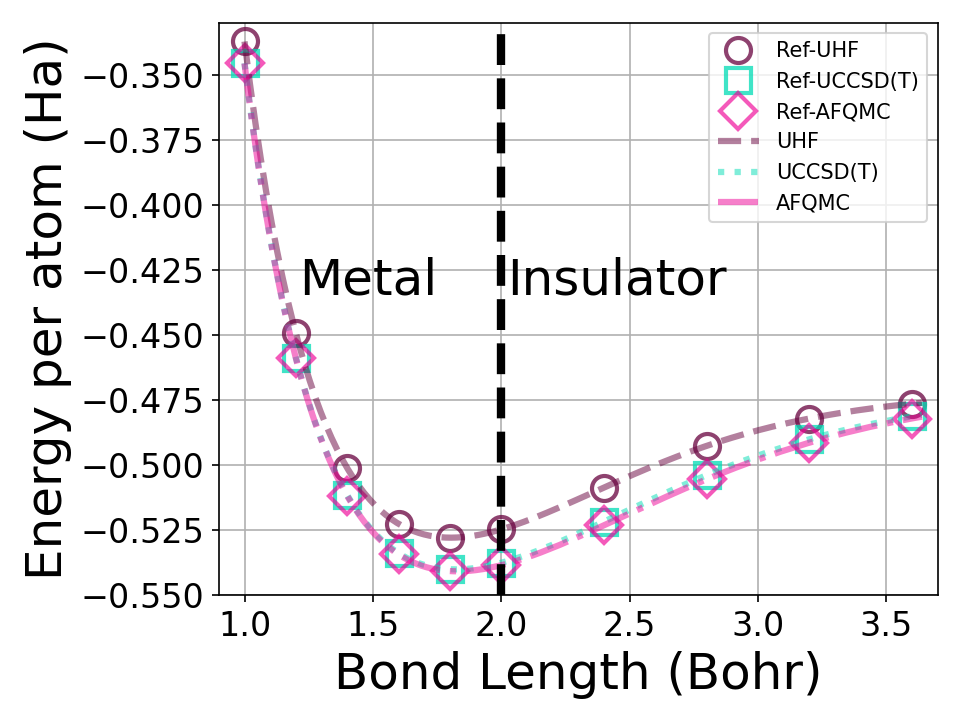}
\caption{Energy per atom vs. bond length for a 50-atom hydrogen chain using the UHF, UCCSD(T), and AFQMC methods in the STO-6G basis. The symbols depict the energies from calculations from Reference (`Ref') \onlinecite{HC1}, while the dotted lines interpolate among 250 of our database energies. AFQMC error bars are too small to see.}
\label{fig:1}
\end{figure}

To analyze the ability of our GPRs to predict the energies of solids in their thermodynamic limit, we first attempt to predict the finite size effects of linear hydrogen chains (LHC) stretched homogeneously, i.e., with their atoms equally-spaced, and with open boundary conditions. This system is a very well-known benchmark for strong electron correlation because of the multireference character it develops at long bond distances and has therefore been used to test the accuracy of a wide-range of many-body methods.\cite{HC1,HC2,Sinitskiy_JCP_2010,Stella_PRB_2011,hachmann2006multireference} As illustrated in Figure \ref{fig:1}, Unrestricted Hartree-Fock theory (UHF) underbinds the hydrogen atoms, while Unrestricted Coupled Cluster Theory (UCCSD(T)) and AFQMC are able to relatively accurately reproduce the chains' energies near their equilibrium bond lengths, but can struggle to capture their energies at longer bond lengths closer to the dissociation limit.\cite{HC1,HC2}

\begin{figure}[h!]
\includegraphics[width=0.48\textwidth]{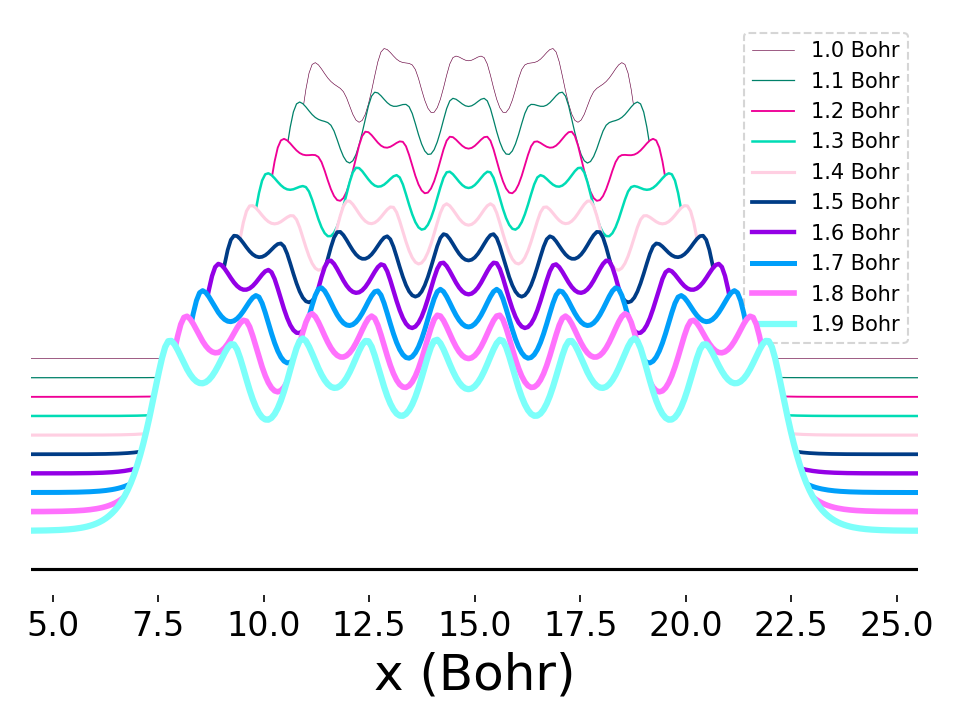}
\caption{Electron density as a function of atomic position (x) for 10-atom hydrogen chains. Each curve depicts the electron density profile when the chain is stretched homogeneously at the bond lengths indicated in the legend. The change in the distance between and depth of adjacent local minima as the bond distance is increased reflects the onset of dimerization. The densities depicted here were computed using Full-Configuration Interaction\cite{szabo} and the y-axis was shifted so that the profiles for all bond lengths could be clearly seen.}
\label{fig:2}
\end{figure}

This system furthermore exhibits a metal-to-insulator transition when stretched homogeneously, which occurs at 1.8 bohr.\cite{HC1} This transition is of second order, meaning that it is continuous with respect to the energy, but can be characterized by the polarization or spin correlation functions.\cite{HC2} Dimerization of pairs of hydrogen atoms in the chains can be observed by looking at the electron density profile along the chains, as in Figure \ref{fig:2}. The maxima correspond to the nuclear positions, while the deep minima indicative of dimerization may be observed between pairs of atoms at all of the chain lengths depicted. 
Methods capable of predicting the energies as a function of bond length must implicitly be able to predict energies across these transitions. 

In order to generate enough data for training, we created a database of the energies of hydrogen chains at varying bond lengths using UHF and two many-body methods - UCCSD(T) \cite{szabo} and AFQMC \cite{afqmc} - in the minimal STO-6G basis. UCCSD(T) has long been considered the gold standard for accuracy for quantum chemistry calculations,\cite{szabo,shavitt_bartlett_2009} and is seeing an increasing number of applications to solids.\cite{McClain_JCTC_2017,Wang_JCTC_2020} AFQMC\cite{afqmc} is a second-quantized QMC method that, despite its typical use of the phaseless approximation,\cite{Zhang_PRL_2003} has been shown to achieve chemical accuracy in systems ranging from small molecules,\cite{Purwanto_JCP_2015,AlSaidi_JCP,Shee_JCTC_2019} to complexes,\cite{Shee_JCTC_2019,Lee_JCTC_2020} to strongly correlated solids.\cite{Ma_PRL_2015,Malone_PRB_2020} As a check on our databases, we produced and extended the benchmarks of Motta \textit{et al.}\cite{HC1} with sub-milliHartree accuracy (see Figure \ref{fig:1}). 
  
To perform our UHF and UCCSD(T) calculations, we use the open source software PySCF.\cite{pyscf} For the AFQMC calculations, we use the high-performance implementation of AFQMC in QMCPACK.\cite{qmcpack} Within QMCPACK, we employ UHF wave functions produced by PySCF as trial wave functions and perform calculations with a time step of $0.005$, 1000 walkers, a Cholesky decomposition threshold of $10^{-8}$, and $10^4$ steps in the phaseless approximation.\cite{afqmc} Energies are computed with the hybrid estimator. Using all of these methods, we compute 250 points for each 10-60-atom chain with bond lengths ranging from 1 to 3.65 bohr. For chains of 70 to 100 atoms, we compute 40 points within the same range of bond lengths in order to conserve computational resources. The 10-30 atom data was used for training, while chains with larger numbers of atoms were used for benchmarking and analysis.\\

\subsubsection{Energy Predictions Using Gaussian Process Regression}

We use the smallest of our hydrogen chains of 10-30 atoms to train and test the GP regressions, which corresponds to 750 total samples. Samples were uniformly mixed by shuffling the data points at all bond lengths for each chain of a given size in the training set. This is to avoid training with an imbalanced data set. SOAP descriptors were constructed by using six GTOs as radial basis functions with a sigma of 1 bohr and six tesseral spherical harmonics as angular functions to build the atomic descriptors for all sizes and bond lengths. A cutoff radius that defines the extent of the atomic environment was set to 7 bohr for all chain sizes and bond lengths. This cutoff radius guarantees that the local environment of an atom consists of a maximum of 14 atoms at the shortest bond lengths studied and a minimum of 2 atoms at the longest bond lengths studied. It may be anticipated that the local atomic environment descriptors become linearly dependent when they have a large cutoff radius and are placed on bulk atoms that repeat throughout the chains. Nonetheless, descriptors placed on the edge atoms manifest asymmetries that reflect the finite extent of the chains. 

The descriptors are first generated for all of the atoms of each chain in the database. A global descriptor is then obtained by averaging each descriptor over the atoms within each chain. Finally, feature selection is carried out by obtaining leverage scores from a CUR decomposition.\cite{cur} The leverage scores are ordered in descending order and features are taken until $97\%$ of the leverage score is accounted for. To perform the CUR decomposition, a Singular Value Decomposition (SVD)\cite{cur} is conducted given a singular value threshold that defines the rank of the decomposition. For this purpose, we used optimal hard thresholding,\cite{optthresh} which makes an optimal choice based on the dimensions and the estimated noise in the features or global descriptors matrix. We don't orthogonalize the features or use covariate principal coordinate analysis to improve our current feature selection, as further discussed in Section \ref{discussion}.\cite{covariates}

A Gaussian kernel with multiple length scales allows more sensitivity to global descriptors without greatly increasing the complexity of the model. We used the maximum likelihood\cite{gprmw} method to optimize the kernel hyper-parameters. We employed up to 500 configurations for training and 250 for validation.

\subsubsection{Accuracy of GPR Predictions} 

After training our GPRs on the UCCSD(T) and AFQMC energies of shorter hydrogen chains, we are able to predict the energies per atom of chains with larger numbers of hydrogen atoms over the same range of bond lengths in the database with reasonable accuracy. We predict the energies per atom using the mean and variance of the posterior distribution. 

\begin{figure*}[ht]
\includegraphics[width=\linewidth]{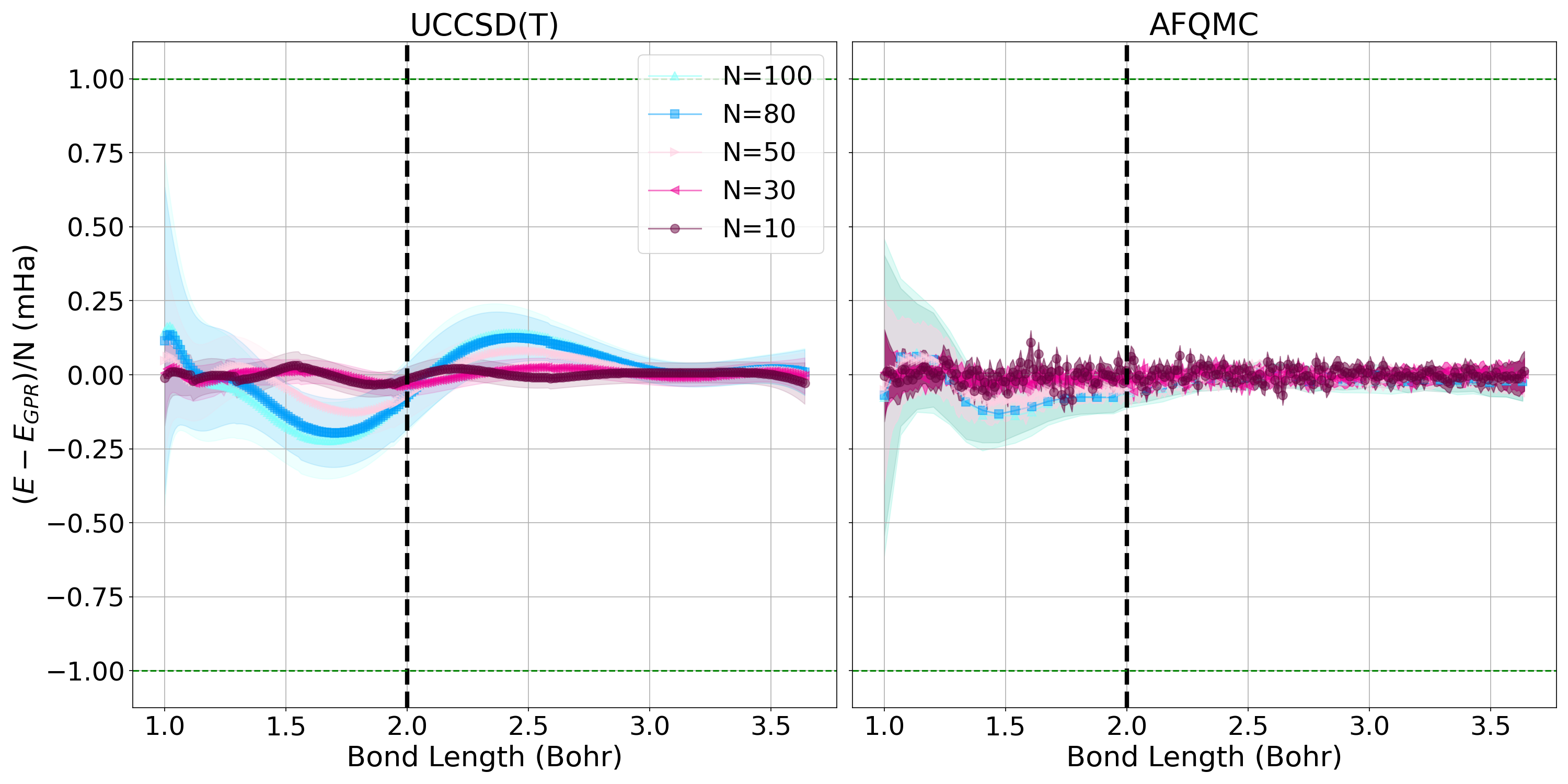}
\caption{Energy differences between the calculated UCCSD(T) (left) and AFQMC (right) energies, and their respective $E_{GPR}$ predictions per atom for hydrogen chains of different lengths in mHa. The green dashed lines depict the bounds of 1 mHa energy differences. The shadows delineate $95\%$ confidence intervals based on the predicted variance. The vertical dashed line denotes the bond length at which the metal-insulator transition occurs.}
6\label{fig:3}
\end{figure*}

Figure \ref{fig:3} depicts the differences between the energies computed with the UCCSD(T) (left) and AFQMC (right) methods, and their respective GPR predictions. In both cases, the differences between the predictions and the calculated energies are less than 1 mHa. It is reassuring to note that the short chain length predictions are most accurate throughout the prediction interval, which is a consequence of training the Gaussian processes on short chains. Prediction errors grow with the lengths of the chains because the generalization error increases with system size. This is reflected in the larger confidence intervals that accompany the larger chain length predictions. 
Hydrogen chains have previously been observed to exhibit slower convergence at short bond lengths because their total chain lengths are not yet long enough to converge finite size effects that stem from long-range Coulomb interactions. This comparatively slow convergence is likely responsible for the larger error bars we observe at short bond lengths.\cite{HC1} At bond lengths longer than 3 Bohr where dissociation begins to occur, the error is significantly smaller and expected to converge faster because chains with longer total lengths will more rapidly converge the long-range Coulomb interaction.

The quality and characteristics of the AFQMC-based GPR predictions are similar to that of the UCCSD(T)-based predictions. Higher accuracies are again observed for shorter chains and at larger bond lengths. The AFQMC-GPR differences are, however, noisier than the UCCSD(T) differences, which reflects the stochastic character of AFQMC. The AFQMC-GPR differences are, in general, smaller than the UCCSD(T)-GPR predictions, especially at intermediate bond lengths. Overall, the AFQMC predictions are slightly more accurate and homogeneous at all of the bond lengths studied, likely due to a larger consistency within the AFQMC data.  

\subsubsection{Extrapolation of Chain Energies to the Thermodynamic Limit}

Given the sub-milliHartree accuracy of these predictions, we now turn to analyzing the performance of our GPR predictions for extrapolating the energies of very long, yet finite chains that approach the thermodynamic limit. In previous studies,\cite{HC1} thermodynamic limit predictions were made by assuming the chain energies varied polynomially with $N^{-1}$, with orders ranging from 1 to 3 depending upon the convergence speed exhibited by the data.\cite{HC1} To make use of such scaling laws, a polynomial must be fit to a large enough number of different chain sizes to capture the correct scaling behavior. To compare the performance of our GP regressions against this more conventional fitting procedure, we fit the energies of chains containing 10, 30, and 50 atoms, as was done in Reference \onlinecite{HC1}. We contrasted the extrapolations produced by this polynomial fit with GPR results trained \textit{once} across different bond lengths on chains of 10, 20, and 30 atoms. 
Indeed, the primary advantage of our method is that we can automatically predict the energy per atom of any chain by computing its global descriptor vector and using the posterior to predict its energy. As an added benefit, the confidence intervals based on the predicted variance provide an estimate of the uncertainty of the prediction, which is not available from typical polynomial regressions. 

\begin{figure*}[htb!]
\includegraphics[width=.95\textwidth]{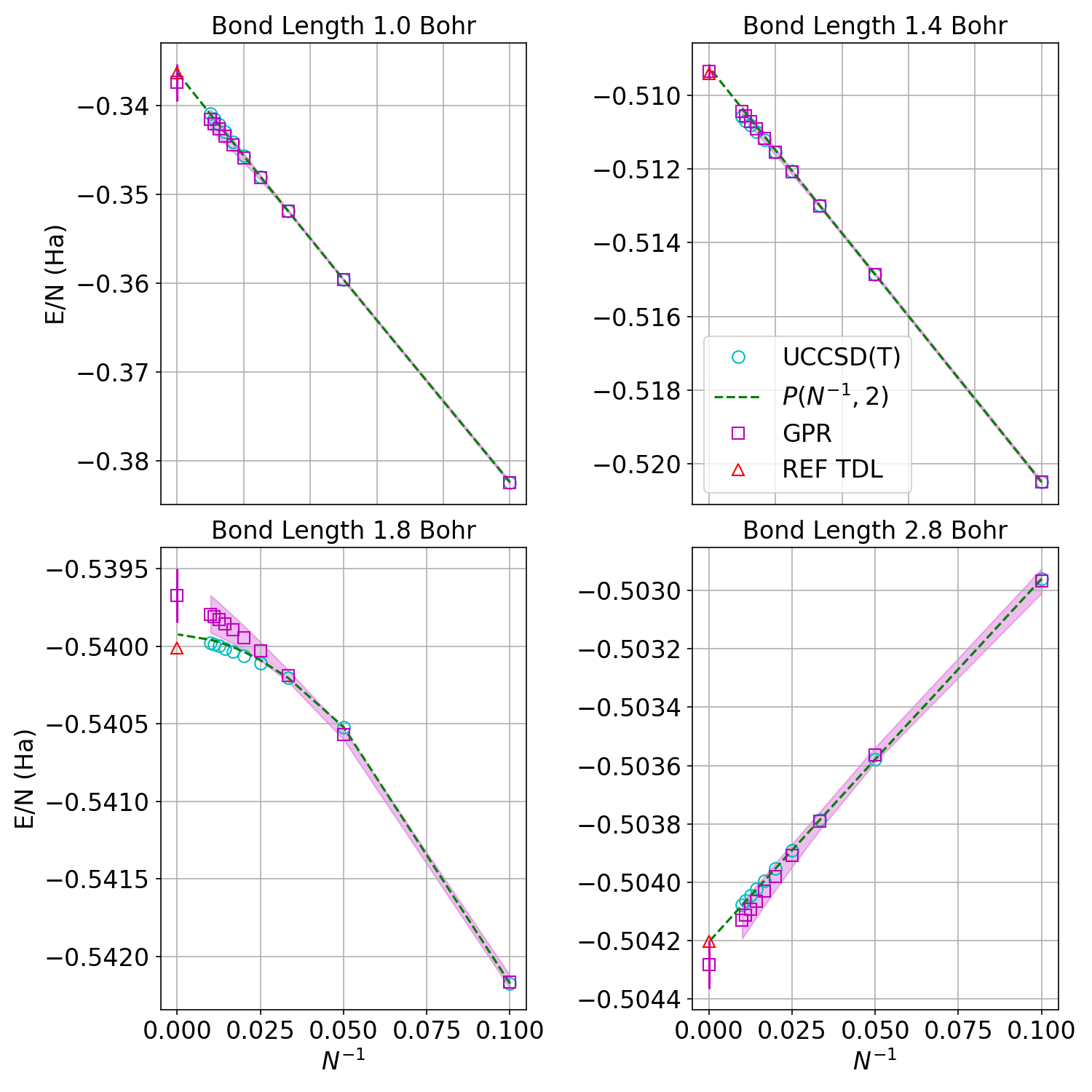}
\label{fig4}
\caption{Predictions of the energy per atom in the thermodynamic limit vs. $N^{-1}$ based on UCCSD(T) results for hydrogen chains with bond lengths of 1.0, 1.4, 1.8, and 2.8 Bohr. Cyan circles denote UCCSD(T) calculations, while maroon squares denote the GPR predictions. The shadow depicts $95\%$ confidence intervals, and the dashed lines depict the polynomial regression of second order at each bond length. The triangle represents the energy in the thermodynamic limit computed in Reference \onlinecite{HC1}. The gap between points at small values of $N^{-1}$ corresponds to chains between 100 and 5000 atoms, which are prohibitive to model even using less expensive theories.}
\end{figure*}

Figure \BR{4} displays the convergence of the energy per atom to the thermodynamic limit for four representative bond lengths. The circles denote the UCCSD(T) calculations while the squares represent the GPR predictions at each size. As before, the shadows delineate $95\%$ confidence intervals on the GPR calculations. The green dashed line denotes the polynomial regression at the given bond length and the red triangle represents the energy in the thermodynamic limit taken from Reference \onlinecite{HC1}. The GPR prediction of the energy in the thermodynamic limit is made using a chain of 5000 hydrogen atoms. As an illustration of the speed of our regression technique, producing the descriptors for the 5000-atom chain took about 2 minutes on an Intel Core i7-8550U (Turbo 4.0 GHz, 4 Cores, 8 Threads) laptop. We note that the differences between the thermodynamic limit predictions made by the reference regression\cite{HC1} and the polynomial regression performed on our dataset simply reflect the small differences between the two different databases. The GPR predictions are in good agreement with the reference and polynomial regressions, deviating most for bond lengths near the equilibrium bond length (around 1.8 bohr) where the convergence is less linear. Note that the energies converge one to two orders of magnitude more rapidly at larger bond lengths because the long-range Coulomb interaction is weaker at larger bond lengths, as described earlier.\\ 

\begin{figure*}[htb!]
\includegraphics[width=0.95\textwidth]{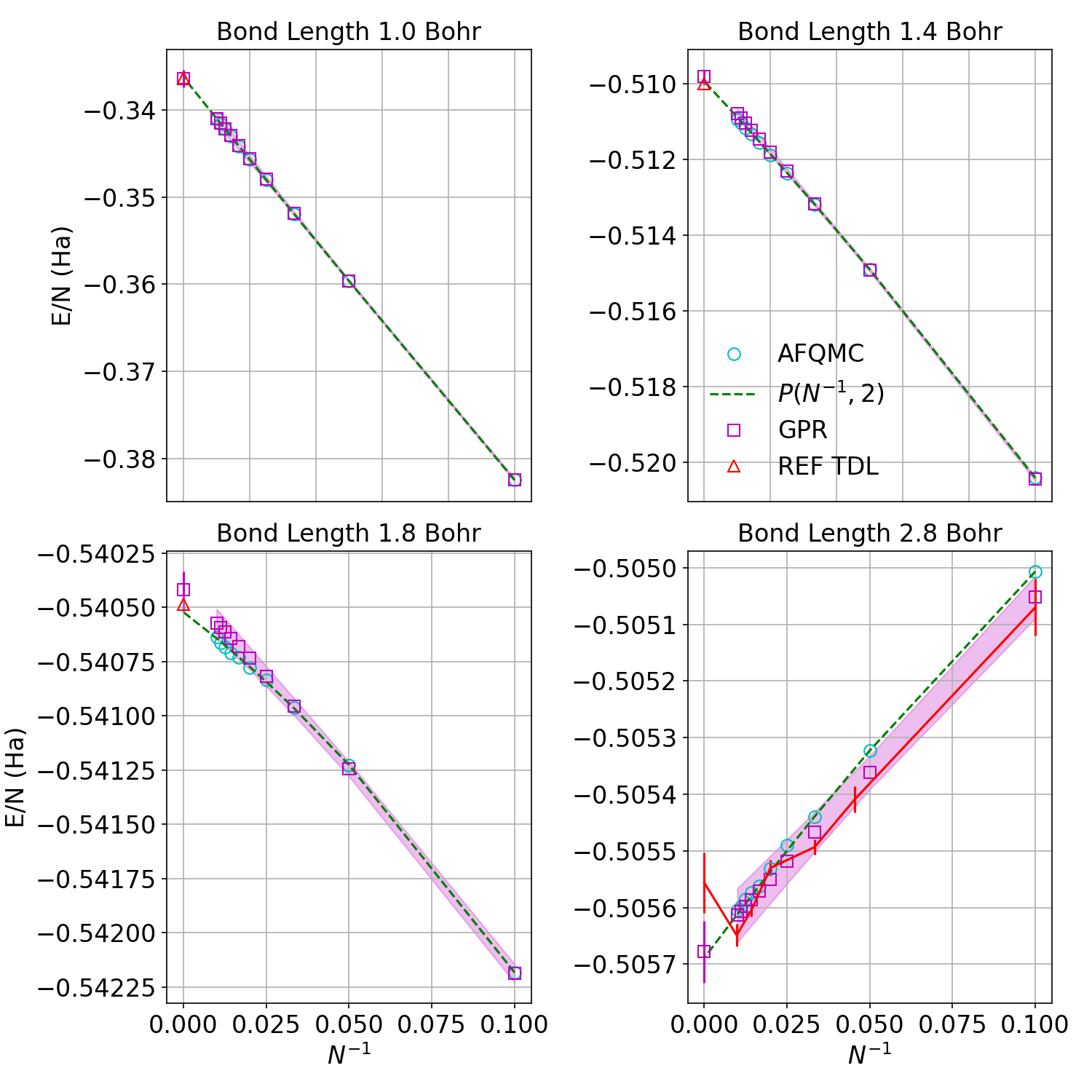}
\caption{Predictions of the energy per atom in the thermodynamic limit vs. $N^{-1}$ based on AFQMC results for hydrogen chains with bond lengths of 1.0, 1.4, 1.8, and 2.8 Bohr. Cyan circles denote direct AFQMC calculations, while the maroon squares denote the GPR predictions. The shadow depicts $95\%$ confidence intervals, and the dashed lines depict the polynomial regression of second order at each bond length. The triangle represents the energy in the thermodynamic limit computed in Reference \onlinecite{HC1}. The gap between points at small values of $N^{-1}$ corresponds to chains between 100 and 5000 atoms, which are too prohibitive to compute using even less expensive theories. Note that the TDL of Reference \onlinecite{HC1} (REF TDL) for the bond length of 2.8 Bohr was replaced by our TDL extrapolation using a polynomial regression because the reference value seemed to be in disagreement with the rest of the reference's data at that bond length.}
    \label{fig:5}
\end{figure*}

Figure \ref{fig:5} similarly exhibits the convergence to the thermodynamic limit for the AFQMC database and its respective GPR predictions. One of the most noticeable differences relative to the UCCSD(T) calculations is that the AFQMC predictions seem more linear close to the equilibrium bond length. This means that the AFQMC calculations can more accurately resolve small, sub-milliHartree differences in the energies as a function of system size and therefore so can the AFQMC-based GPR.\\

\begin{figure*}[htb!]
\includegraphics[width=0.95\textwidth]{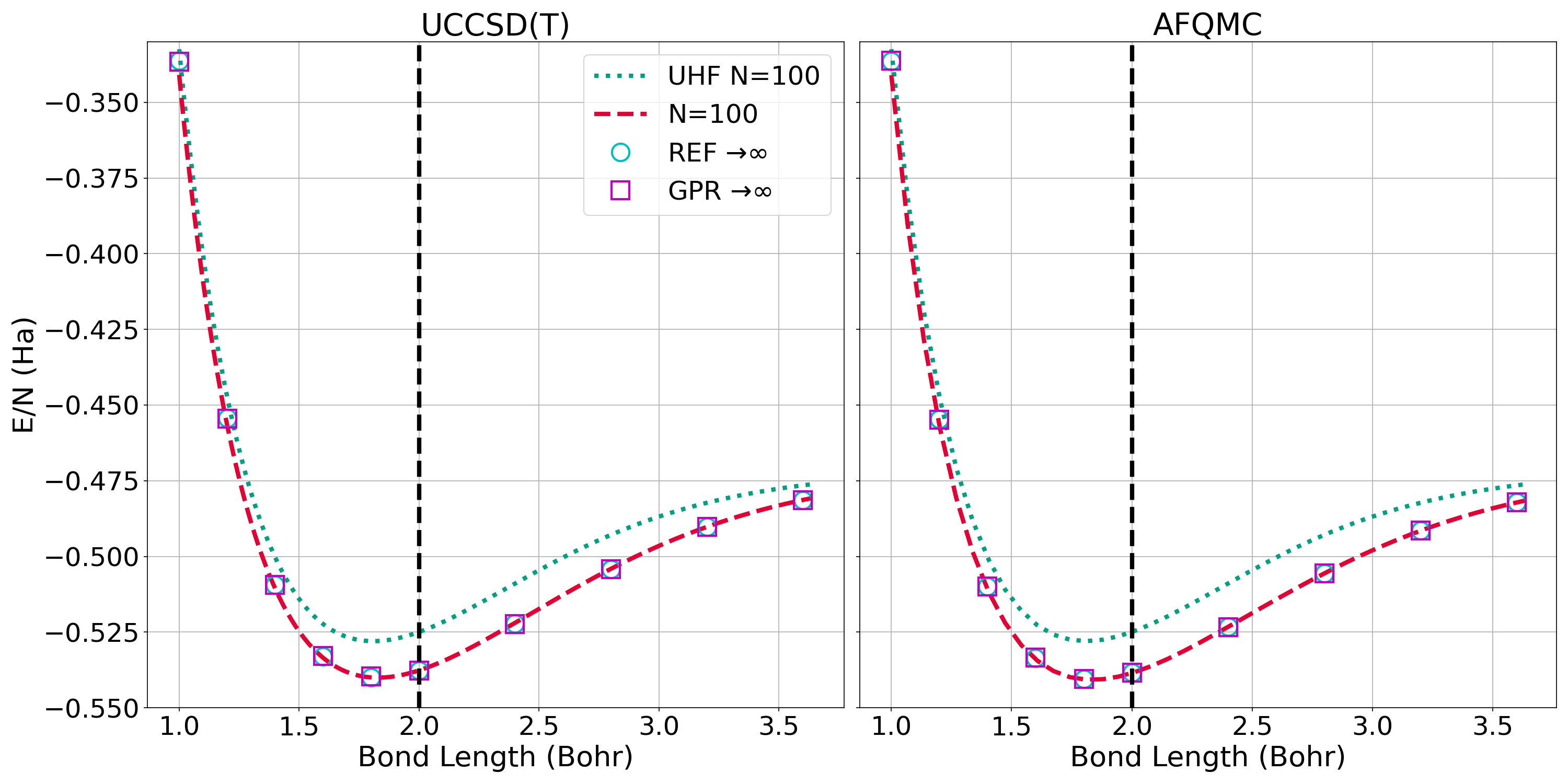}
\caption{Energy per atom computed for $N=100$ chains and predicted for $N=\infty$ chains using the UCCSD(T) (left) and AFQMC (right) methods. The ``REF $\to\infty$" is the TDL extrapolation taken from Ref. \onlinecite{HC1}. We plot this reference's extrapolation so that it can be contrasted with our GPR's prediction using 5000 atoms.}
    \label{fig:6}
\end{figure*}

The left panel of Figure \ref{fig:6} presents the energy of the hydrogen chains as a function of bond length directly calculated using UHF and UCCSD(T) for 100-atom chains, as well as the reference\cite{HC1} and GPR predictions in the thermodynamic limit. The energy differences between the $N=100$ UCCSD(T), reference, and GPR predictions are hardly perceptible. The right panel of Figure \ref{fig:6} likewise presents the energy as a function of bond length for the largest, $N=100$-atom AFQMC calculations we were able to perform, in addition to the reference and GPR thermodynamic limit predictions. As in the UCCSD(T) case, the discrepancies are too small to discern at this scale.\\ 

\begin{figure*}[ht!]
\includegraphics[width=0.95\textwidth]{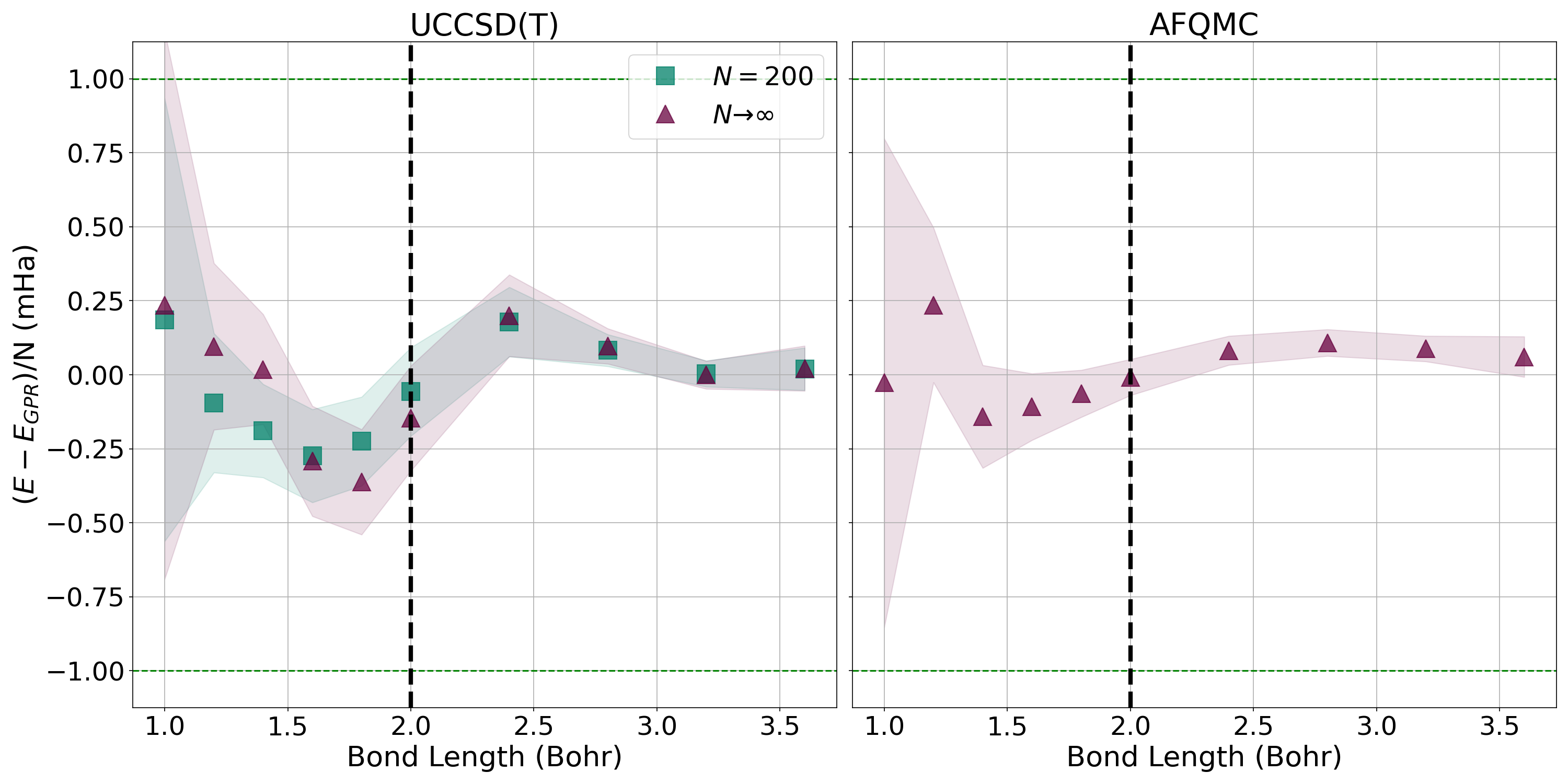}
\caption{Differences between the energies predicted by Ref. \onlinecite{HC1} and the UCCSD(T) (left) and AFQMC (right) GPR-predicted energies in the thermodynamic limit (red triangles). For both the UCCSD(T) and AFQMC plots, we assume that the GPR prediction using 5000 atoms is representative of the GPR prediction in the thermodynamic limit. On the left, we also plot the UCCSD(T) energies for $N=200$ hydrogen chains, the largest we could directly simulate. The shadows depict $95\%$ confidence intervals for the GPR predictions. }
    \label{fig:7}
\end{figure*}

To more closely examine how the GPR predictions converge with the number of atoms in the chains, in Figure \ref{fig:7}, we plot the  difference between the thermodynamic limit predictions of Ref. \onlinecite{HC1} and our UCCSD(T) (left) and AFQMC-based (right) GPR predictions on the milliHartree scale. For both methods, we take $N=5000$ hydrogen chain GPR predictions to be representative of the thermodynamic limit. In the left-hand panel, we also plot UCCSD(T) results for $N=200$ hydrogen chains, the largest we could directly simulate, to contrast $N=200$ with $N\rightarrow\infty$ results. We see that, at smaller bond lengths, discrepancies still remain between the $N=200$ and $N\rightarrow \infty$ results, signifying that finite size effects still influence the energies of even $N=200$-length chains.\\

These discrepancies are also manifested in the larger confidence intervals that accompany the GPR predictions. Even so, GPR predictions at all bond lengths studied possess sub-milliHartree accuracy, and the discrepancies between the different chain length predictions disappear at the longest bond lengths studied. In contrast, the right panel of Figure \ref{fig:7} demonstrates that the AFQMC-based GPR predictions are in much better agreement with Ref. \onlinecite{HC1}'s thermodynamic limit predictions, even at shorter bond lengths. This is in line with the results presented earlier in Figure \ref{fig:3}.\\ 

\begin{figure*}[ht!]
\includegraphics[width=0.95\textwidth]{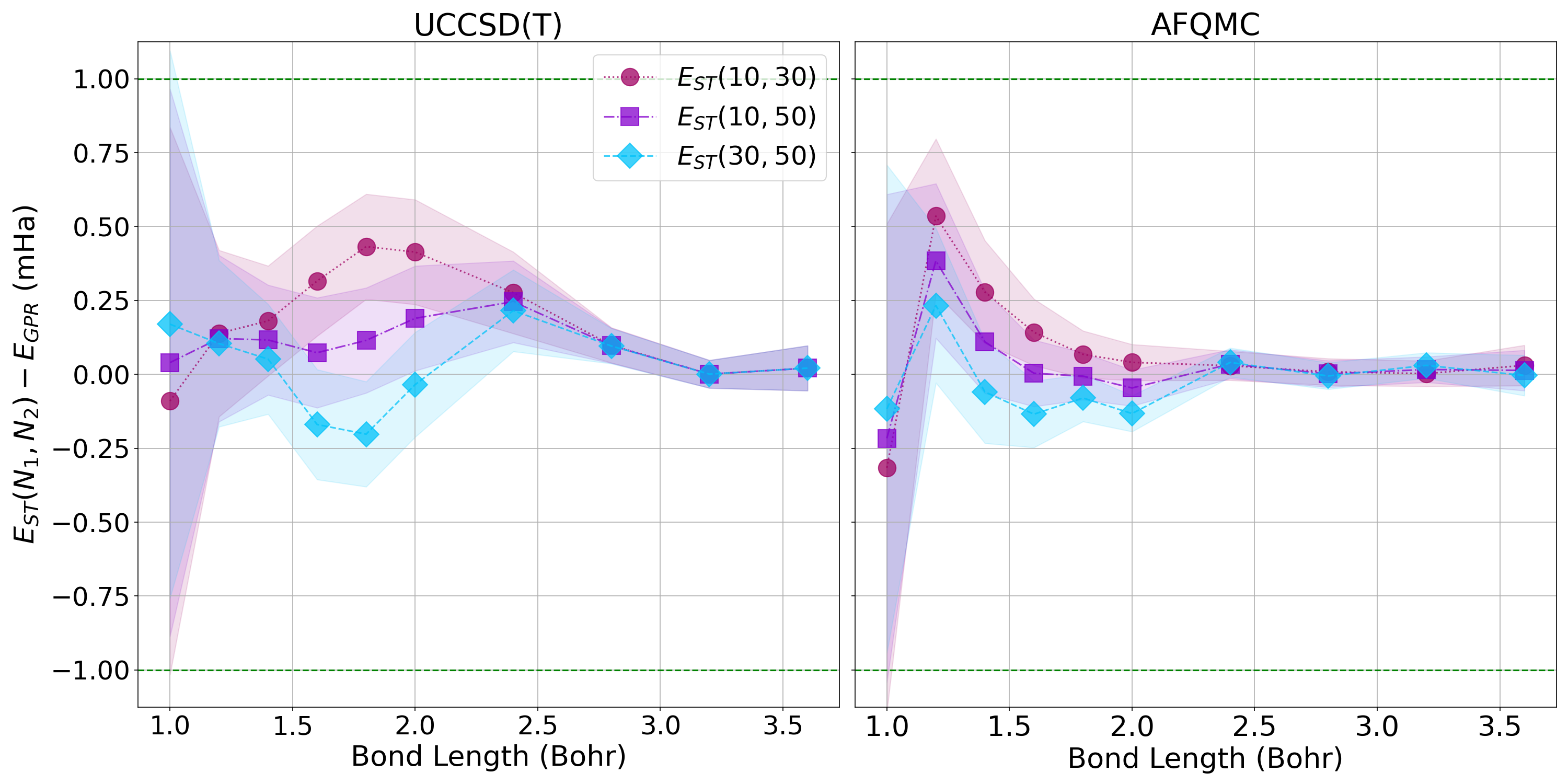}
\caption{The difference in energies between our GPR predictions in the thermodynamic limit, $E_{GPR}(N\rightarrow\infty)$, and extrapolated energies obtained using the subtraction trick, $E_{ST}$. (Left) Differences based upon UCCSD(T) energies; (Right) differences based upon AFQMC energies.}
\label{fig:8}
\end{figure*}

Since our calculations were performed with open boundary conditions (OBC), it is also worthwhile to compare our predictions to those produced using the ``subtraction trick,''\cite{subtrick} in which the energies of systems of different sizes are subtracted to eliminate surface effects from bulk energies. 
Figure \ref{fig:8} presents the differences in energy between our GPR predictions of the energies in the thermodynamic limit and those produced using the subtraction trick based on chains of different lengths. The differences in the energies predicted by these approaches is sub-milliHartree at all bond lengths studied, further demonstrating that our GPR predictions are highly accurate relative to a widely-employed benchmark, while also illustrating the surprising accuracy of the subtraction trick. As the subtraction trick eliminates edge effects from energy predictions, this comparison especially highlights the GPR method's ability to correct for edge effects. It is satisfying to see that the energies predicted by the subtraction trick performed on chains of lengths 30 and 50, which should yield the most accurate predictions of the subtraction trick calculations, are in the greatest agreement with our GPR predictions, especially at intermediate bond lengths. As before, we see that our GPR predictions are in the greatest agreement with the subtraction trick results at longer bond lengths. Indeed, our GPR predictions almost perfectly agree with all three of the subtraction trick predictions at the longest bond lengths studied. Moreover, our AFQMC-based GPR predictions again converge more rapidly and reliably to the thermodynamic limit with increasing bond length. Overall, Figures \ref{fig:7} and \ref{fig:8} possess very similar features: the GPR predictions overestimate the energies at the shortest bond lengths and then oscillate between under- and overestimating the energies at intermediate bond lengths before coming to agreement at the longest bond lengths. This points to the overwhelming agreement between the polynomial regression and subtraction trick energies. These comparisons also demonstrate that the GPR predictions are not uniformly biased toward over- or underestimating energies. 

A more quantitative comparison of the predictions generated by all of these methods can be found in the Supplementary Materials.  

\subsection{Two-Dimensional, Inhomogeneous Hydrogen Chains}

\begin{figure*}[ht!]
\includegraphics[width=0.65\textwidth]{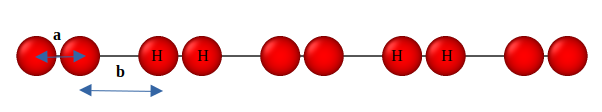}
\caption{Illustration of the linear chains of hydrogen dimers studied in this work with intradimer distance, $a$, and interdimer distance, $b$.}
\label{fig:lhd}
\end{figure*} 

Given the success of GPR at predicting the energies of homogeneously-stretched hydrogen chains in the thermodynamic limit, we next examine the capacity for the same GPR techniques to predict the energies of inherently heterogeneous chains of hydrogen dimers. As depicted in Figure \ref{fig:lhd}, these chains of hydrogen dimers are described by two key distances: the intra-dimer bond distance, $a$, and the inter-dimer bond distance, $b$. In the following, we generally fix the intradimer distance, $a$, between 1.0 and 3.5 bohr, and vary the interdimer distance between 1.0 and $a$ bohr, maintaining open boundary conditions. While these chains of dimers enable us to retain the same periodicity present in our earlier homogeneous chains, they also enable us to purposefully and controllably introduce heterogeneity into our systems that complicates our prediction problem. Indeed, these chains of dimers manifest several levels of correlation when stretched, typically necessitating the use of advanced quantum chemistry methods to make high-accuracy energy predictions.\cite{dimmers2007} 

\begin{figure*}[ht!]
\includegraphics[width=1.0\textwidth]{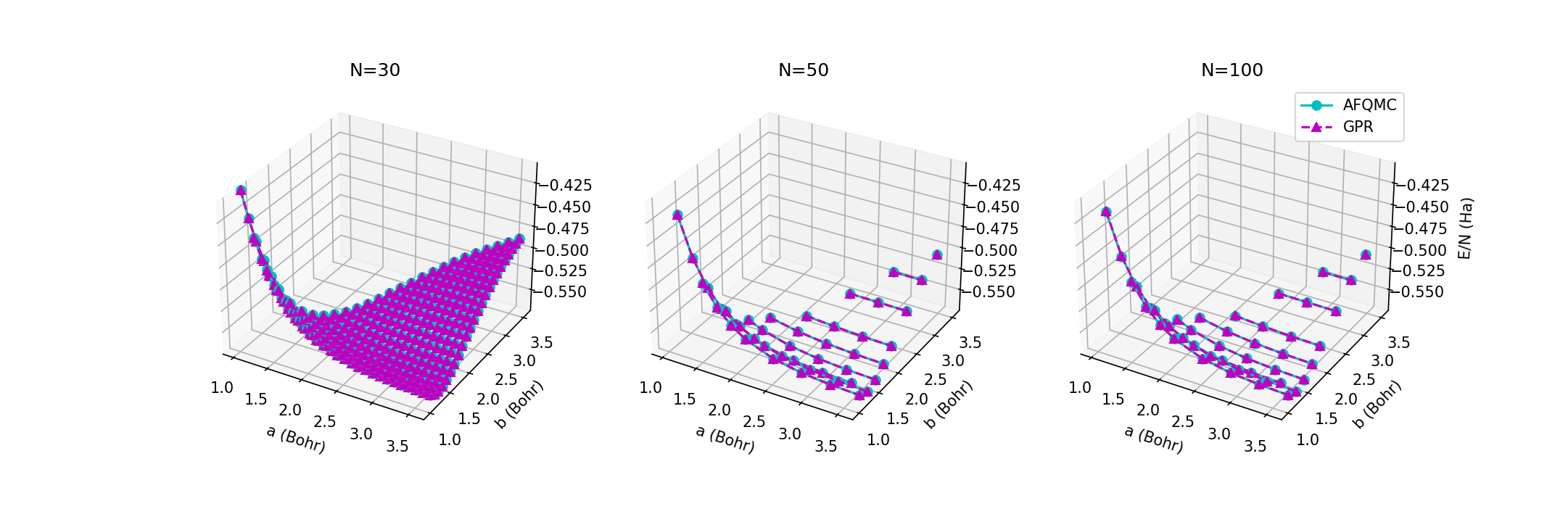}
\caption{Energy surfaces, $E/N$, predicted for chains consisting of (left) 30, (center) 50, and (right) 100 atoms (15, 25, and 50 dimers, respectively) for several $a$ and $b$ values. AFQMC energies are given by the cyan circles, while GPR predictions are given by the maroon triangles.}
\label{fig:3dgrid}
\end{figure*} 

To study the performance of our GPR algorithm on these chains, we generate a database of dimer chain energies staring from UHF calculations with single Slater determinants that we again input into either CCSD(T) or AFQMC calculations. We model our hydrogen atoms using the minimal STO-6G basis set given the steep computational cost of the system with increasing system size. Chains of 5, 10, and 15 dimers for a total of 176 configurations  were employed for training. The remaining 315 configurations of chains consisting of 20 to 50 dimers were subsequently used for testing and validation. The same atomic environment descriptors previously employed for the homogeneous chains were also employed here.  

\begin{figure*}[ht!]
\includegraphics[width=0.95\textwidth]{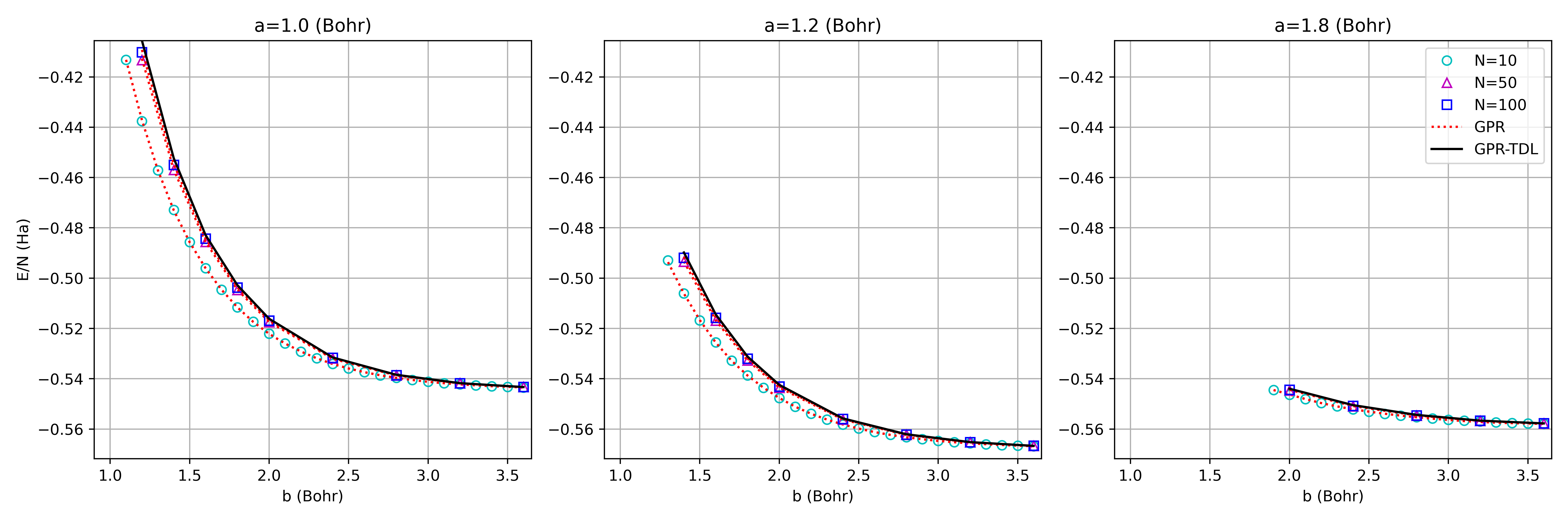}
\caption{Comparison of GPR and AFQMC predictions for different intra-dimer bond lengths as a function of inter-dimer distances. The $N$=10, 50, and 100-atom data are all provided by AFQMC.}
\label{fig:2dab}
\end{figure*}

The energy surfaces for chains consisting of $N=30$, $50$, and $100$ atoms are depicted in Figure \ref{fig:3dgrid}. It can be seen that the GPR predictions are in qualitative agreement with the AFQMC database values, both for short chains ($N$=10) and long chains approaching the thermodynamic limit ($N$=100). In particular, GPR is able to well describe both the energy minimum around $a$=1.5 bohr, $b$=3.25 bohr, and highly stretched chains with both $a$ and $b$ greater than 3 bohr. More detailed slices of the potential energy surface for several values of $a$ are depicted in Figure \ref{fig:2dab}. 

As is apparent from these plots, the approach to large inter-dimer separations is highly dependent upon the intra-dimer separation: for small $a$, the approach is steeper than for large $a$. This behavior is a sign of correlation between the $a$ and $b$ values and is non-trivial, given the seeming simplicity of the model. This makes the model a useful testbed for multidimensional extrapolations, as further discussed in the Supplementary Materials.

\begin{figure*}[ht!]
\includegraphics[width=0.98\textwidth]{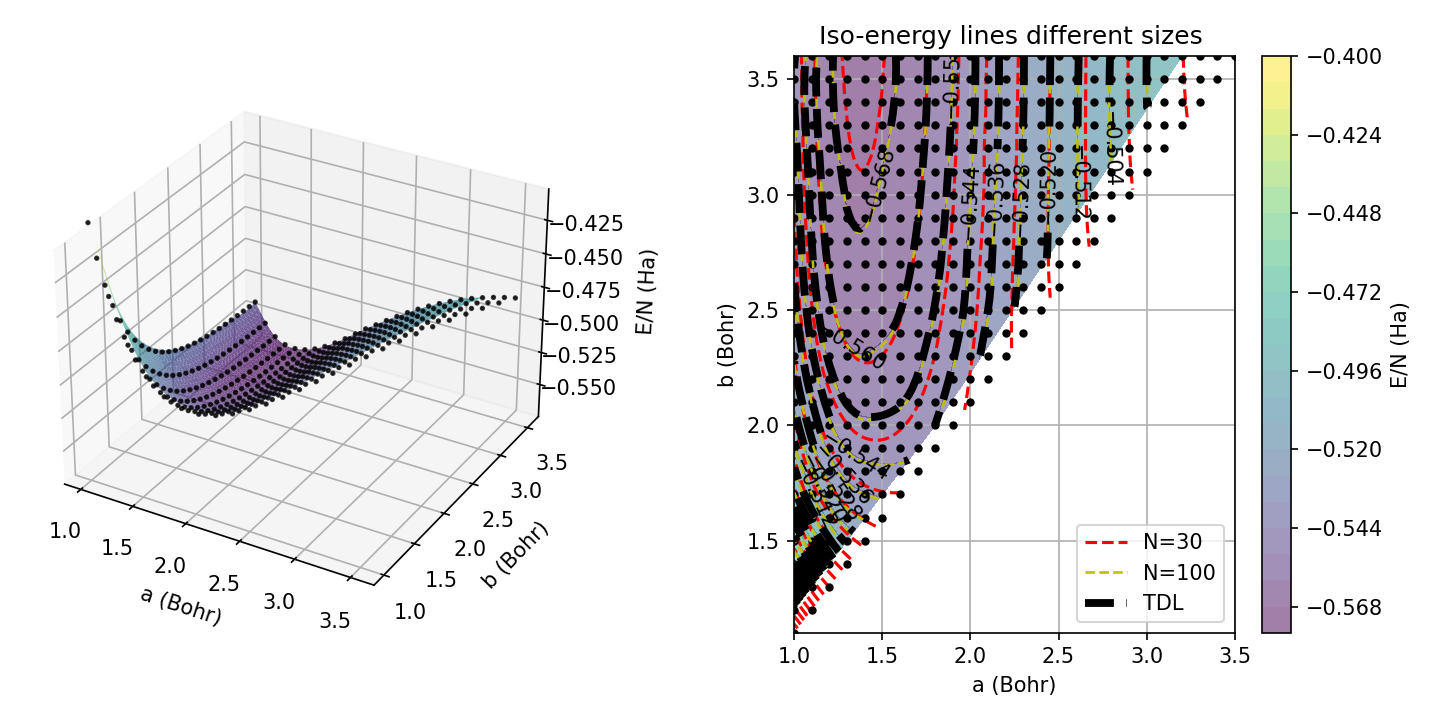}
\caption{(Left) Interpolated GPR potential energy surface, $E/N$, for a chain of 15 dimers as a function of $a$ and $b$. The black dots denote the AFQMC training points from the database. (Right) Iso-contours of the energy surface, $E/N$, interpolated for chains of $N=30$ (red) and $N=100$ (yellow) atoms, and in the thermodynamic limit (dashed black lines). The color map denotes the GPR predictions of the energy in the thermodynamic limit.}
\label{fig:2d_tdl}
\end{figure*}

To visualize the energy surface, as shown in the Figure \ref{fig:2d_tdl}, we use triangulation over the sample points and then Delaunay smoothing. The thermodynamic limit was estimated using GPR regression on $N$=5000 atoms. On the left, we present the 3D energy surface for a chain of 15 dimers; the black dots denote the energies estimated by GPR in the thermodynamic limit. On the right, we provide a heat map corresponding to the plot on the left annotated with iso-energy contour lines predicted using GPR for systems of different sizes. In particular, the red and yellow dashed lines denote the energies for chains comprised of 15 and 50 dimers, respectively. The errors on these energies are all less than 1 mHa, which is within chemical accuracy.

This plot underscores how the contours change or shift with system size. We can see that the largest differences between the contours occur near the minimum of the plot around an intra-dimer distance of 1.5 bohr and an interdimer distance of 3 bohr. In this region of the surface, the $N=30$ contours differ significantly from the $N=100$ contours, which nearly align with the thermodynamic limit contours, suggesting that 100 atoms are nearly enough to converge simulations of this system to their TDL. Away from this minimum, the contours for all three system sizes concur, demonstrating that the system experiences weaker finite size effects for these parameters. GPR's success extrapolating the energies of this non-trivial, multidimensional model suggest that it is likely to have similar success on the more complex models and solids of interest to the wider scientific community.

\section{Discussion of Results \label{discussion}} 

Although we employed Gaussian Process Regression in this work, a wide range of other machine learning approaches, including artificial neural networks, could also be used to perform these extrapolations. We opted to employ kernel methods like Gaussian processes because they are non-parametric and make use of Bayesian inference at a comparatively low, $O(N_t^3)$ cost, where $N_t$ is the size of the training set.\cite{metha} It has been proposed\cite{CuiKrems} as a rule of thumb to use $N_t=10\times d$, where $d$ is the dimension of the feature space, to train a GPR. In contrast, neural network-based approaches involve matrix-vector multiplications that scale with the number of neurons in the network, $N_{n}$, and the dimension of the input vector, $d$, as $O(N_n d)$. If the number of neurons in the network is small, this implies that neural networks are less expensive to employ than GPR. However, neural networks typically necessitate the use of non-linear activation functions that may increase their overall cost. More importantly, neural networks often suffer from overfitting if care is not taken to reoptimize their number of nodes or structures. Overfitting is much less of a concern for GPR since GPR with the same kernel but more training points is guaranteed to be more accurate. In practice, NNs use at least 2 to 3 orders of magnitude more training data than GPR.\cite{metha,NNvsGPR} When training data is scarce - as it usually is when many-body electronic structure calculations are involved - GPR-based techniques hence become the method of choice.\cite{NNvsGPR} One may also ask whether using GPR on these low-dimensional data sets is more sophisticated than necessary and if other, less sophisticated regression techniques based on a small number of parameters could instead be employed. As demonstrated in the Supplementary Material, we have compared the performance of our GPR approach to that of Bayesian Multivariate Adaptive Regression Splines, a spline-based technique, and found that our GPR approach can extrapolate with significantly greater accuracy. We moreover show that, while one can extrapolate using a few simple parameters, this extrapolation is not readily generalizable to more complex situations in which the parameters to use are less obvious. Lastly, as illustrated throughout this manuscript, GPR inherently quantifies uncertainties, which are critical for being able to determine its accuracy relative to that of other methods. 

Our studies of low-dimensional hydrogen chains naturally beg the question of how well our techniques can be generalized to more realistic multidimensional solids that are accompanied by an even more rapid growth in computational expense.
Much like other GAP methods, our approach should be readily generalizable to higher dimensional systems, given sufficient data and high-quality features. Indeed, here, we took the first step toward demonstrating this by applying our model to both a one-dimensional and a nontrivial two-dimensional system, and in a previous preprint, we demonstrated how a similar GPR-based approach could be leveraged to predict the energies of 3D alloys.\cite{Borda} The key challenge associated with higher-dimensional predictions is the curse of dimensionality: the higher the dimensionality of the space, the more data that is needed for training to learn the larger space with sufficient accuracy to make effective comparisons between different atomic environments. The resulting increase in cost can be slowed through a more judicious selection of features and design of kernels. CUR\cite{cur} decompositions and Kernelized Principal Covariates Regression\cite{covariates} are excellent alternatives for identifying the most relevant features, which can significantly reduce the dimension of the descriptors of a given data set. More effective kernels may also be constructed through approaches that recursively evaluate the differences between structures.\cite{ceriotti} For example, De \textit{et al.} proposed kernels based on regularized structure matching to optimize the comparisons between the atomic environments of different structures.\cite{ceriotti}  Thus, with further technical developments, we believe that the techniques presented here should be readily generalizable to the even larger, more complicated solids that they would most benefit.

\section{Conclusions \label{conclusion} }
In summary, in this work, we have presented a Gaussian Process Regression-based approach for predicting the many-body energies of hydrogen chains, the simplest examples of \textit{ab initio} solids, in the thermodynamic limit. We have shown that, by training on databases of the energies of short (10-30-atom) homogeneous and inhomogeneous hydrogen chains with varying intra- and inter-dimer distances, we can predict the energies of these chains in the thermodynamic limit with sub-milliHartree accuracy relative to predictions made by alternative extrapolation techniques. These alternative techniques, including polynomial regressions and the ``subtraction trick,'' typically necessitate computing the energies of chains much longer than the chains employed in our training sets. As such, our approach enables the highly accurate prediction of the energies of solids in the thermodynamic limit based upon relatively small systems, and hence, much less expensive calculations. Unlike many finite size extrapolation techniques which apply to systems with only certain geometries, densities, and/or dimensionality, as demonstrated by the easy generalizability of our method to both homogeneous and inhomogeneous chains, our approach is largely agnostic to the physical characteristics of the system studied; as long as there is sufficient and representative training data, our approach can be applied, making it particularly useful for some of the more complex systems of modern interest, such as those at interfaces or having irregular geometries.  

\section{Data Availability}
The data that support the findings of this study are openly available online at \url{https://github.com/josuelandinez/LHC_Database}.

\section{Supplementary Material}

A supplementary materials document has been provided that contains comparisons of the accuracy of our GPR methods to those of other regression techniques and tables of the energies predicted by the different approaches described here that underlie many of the plots provided in the main text.

\section{Acknowledgements}
E.J.L.B. would like to thank Amit Samanta for insightful discussions and Qiming Sun for his kind assistance with performing calculations in PySCF. E.J.L.B. (concept, modeling, analysis, manuscript preparation), A.L. (manuscript preparation), and B.R. (concept, mentoring, manuscript preparation) graciously acknowledge support from U.S. Department of Energy, Office of Science, Basic Energy Sciences Award \#DE-FOA-0001912. B.R. (concept) also received support from U.S. National Science Foundation grant OIA-1921199 and the Research Corporation of America (concept). This research was conducted using computational resources and services at the Center for Computation and Visualization, Brown University. 

\section{References}
\bibliographystyle{unsrtnat}
\bibliography{biblio2}

\end{document}